\documentclass[12pt]{iopart}

\usepackage{graphicx}
\usepackage{iopams}

\usepackage{epsfig}
\usepackage{psfrag}
\usepackage{amssymb}
\usepackage[usenames]{color}

\newlength{\figwidth}
\newcommand{\units}[1]{\,\mathrm{#1}}
\newcommand{\parref}[1]{(\ref{#1})}

\newcommand{\dexp}[1]{\cdot 10^{#1}}

\newcommand{\Ox}{\mathrm{O}}
\newcommand{\Ni}{\mathrm{N}}

\begin{document}
\title{Positive and negative streamers in ambient air: modeling evolution and velocities}
\author{Alejandro Luque$^1$, Valeria Ratushnaya$^1$ and Ute Ebert$^{1,2}$}
\address{$^1$~CWI, P.O. Box 94079, 1090 GB Amsterdam, The Netherlands
\\$^2$~Department of Physics, Eindhoven University of Technology,
  The Netherlands}

\begin{abstract}
We simulate short positive and negative streamers in air at standard temperature and pressure. They evolve
in homogeneous electric fields or emerge from needle electrodes with voltages of 10 to 20~kV. The streamer
velocity at given streamer length depends only weakly on the initial ionization seed, except in the case of
negative streamers in homogeneous fields. We characterize the streamers by length, head radius, head charge
and field enhancement. We show that the velocity of positive streamers is mainly determined by their radius and
in quantitative agreement with recent experimental results both for radius and velocity. The velocity of negative
streamers is dominated by electron drift in the enhanced field; in the low local fields of the present
simulations, it is little influenced by photo-ionization. Though negative streamer fronts always move at
least with the electron drift velocity in the local field, this drift motion broadens the streamer head,
decreases the field enhancement and ultimately leads to slower propagation or even extinction of the
negative streamer.
\end{abstract}


\submitto{cluster issue on ``Streamers, Sprites and Lightning'' in \JPD}

\maketitle

\section{Introduction}

Atmospheric pressure corona discharges are widely used in technology. Streamers, which are the basic
building blocks of these discharges, focus a large part of the energy of the reactor into a small volume. As
positive streamers emerge from pointed electrodes at lower voltages than negative
ones~\cite{Raizer91,BrielsPM}, recent investigations have largely focused on positive streamers. However,
negative streamers are clearly present in many natural phenomena of atmospheric electricity like lightning,
sprite discharges etc.~\cite{Raizer91,Baz,Pasko2002/GeoRL,Williams06/PSST}. Furthermore, modern high voltage
supplies easily create negative streamers~\cite{BrielsPM,Winands2006/JPhD,Winands2008/JPhD}, and they are
very promising for disinfection applications~\cite{win07} if electrical matching problems can be overcome.
An experimental study of positive and negative streamers in air at standard temperature and pressure in a
wide voltage range is available in~\cite{BrielsPM}.

The simulation of positive streamers in three spatial dimensions with cylindrical symmetry meanwhile is
based on a large body of research. Pioneering work was done by Wang and Kunhardt~\cite{Kunhardt} and by
Dhali and Williams \cite{Dhali1987/JAP}. The use of more complete and realistic plasma-chemical models
\cite{Kossyi1992/PSST}, better modeling of the electrode geometry \cite{Babaeva1996/JPhD,
Abdel-Salam2007/JPhD} and an efficient calculation of the non-local photo-ionization source
\cite{Bourdon2007/PSST, Luque2007/ApPhL} have finally allowed simulation and experiment results to converge
within a narrow range.  Pancheshnyi {\it et al.}~\cite{Pancheshnyi2005/PhRvE} were able to predict the mean
streamer velocity at varying pressures within a range of around 25\% and thus question the role of
photo-ionization versus fast electron detachment in repetitive positive streamer discharges in air
\cite{Pancheshnyi2005/PSST}.  Also remarkable was the reproduction
of experimental results \cite{Yi2002} of streamers in long gaps of $13\units{cm}$ performed in \cite{Pancheshnyi2003/JPhD}.

In early work, the non-local ionization mechanism through photons was replaced by background ionization, and
positive and negative streamers looked fairly similar. An example of a simulation of a double-headed
streamer that is completely dominated by the assumption of the initial ionization distribution, can be found
in~\cite{Kuli1994}. Since photo-ionization was introduced as a non-local ionization mechanism in air to
explain the propagation of positive streamers, precisely three groups of authors have investigated negative
streamers in air with photo-ionization. Babaeva and Naidis have compared positive and negative streamers
emerging from pointed electrodes in a short paper in 1997~\cite{Babaeva1997/ITPS}, Liu and Pasko have
investigated doubled-headed streamers in homogeneous fields~\cite{LiuP04,Pasko2007/PSST}, and the
present authors have studied the influence of photo-ionization on propagation~\cite{Luque2007/ApPhL} and
interaction~\cite{Luque2007/arXiv} of streamers of both polarity.

The present paper is devoted to a systematic study, characterization and comparison of positive and negative
streamers in ambient air. It is organized as follows. In section~\ref{sect:model} we describe our model.
Section~\ref{sect:homo} treats double-headed streamers in homogeneous fields, their dependence on the
ionization seed chosen as a starting point for the simulations and their basic mode of propagation.
Section~\ref{sect:inhomo} treats positive and negative streamers emerging form needle electrodes; streamers
of both polarity are characterized by velocity, field enhancement, head radius and head charge;
characteristic differences are found. Their velocities are dominated either by the head radius for positive
streamers or by the enhanced field for negative streamers. Section~\ref{exp} shows a convincing comparison
with the experiments in~\cite{BrielsPM}. Finally, we summarize our main results in
section~\ref{sect:summary}. \ref{CST} contains the Charge Simulation Technique for the needle electrode, and
\ref{defs} the definition of the quantities used for the electro-dynamic characterization of the streamer.

\section{Model}
\label{sect:model}

We use a fluid model of air that contains electrons and six species of ions: $\Ni_2^+$, $\Ox_2^+$,
$\Ni_4^+$, $\Ox_4^+$, $\Ox_2^+\Ni_2$ and $\Ox_2^-$. While electrons diffuse and drift in a self-consistent
electric field, the ions due to their much larger mass can be approximated as immobile. We consider 15
reactions among the species, taken from \cite{Pancheshnyi2005/PhRvE} plus photo-ionization which we
approximate as described in \cite{Luque2007/ApPhL}. All our simulations are performed at standard pressure
and temperature. We assume cylindrical symmetry of the streamer and we solve the model numerically by means
of adaptively refined grids \cite{Montijn2006/JCoPh}.

We performed numerical experiments in two geometries:  the simplest one is defined by plane parallel
electrodes with fixed electrical potential, therefore we impose Dirichlet boundary conditions to the
potential. A more complex geometry is given by a needle electrode, which we simulate by means of a
simplified version of the Charge Simulation Technique, as detailed in \ref{CST}.

To facilitate the physical interpretation of the simulation results we provide, besides some direct results,
a number of derived quantities to characterize the streamer evolution.  The precise definition of these
quantities (length $L$, velocity $v$, enhanced field $E_{max}$, radius $R$ and charge in the streamer head
$Q$) is provided in \ref{defs}. The choice of these particular quantities is guided by the aim to develop an
electro-dynamic model of streamer heads and streamer channels in the future; qualitative models of this type
have been suggested in the past Russian literature~\cite{DyaKach1,DyaKach2,RaizerSim,Baz}. However, these
models require further improvements as was discussed in~\cite{Ebert2006/PSST}.

\section{Double-headed streamers in homogeneous background fields}
\label{sect:homo}

\subsection{Motivation}

We first study streamers in homogeneous background fields. Most streamer experiments are performed in
needle-plane~\cite{Briels2006/JPhD,BrielsScaling,BrielsPM}, wire-plane or wire-cylinder
geometries~\cite{Winands2008/JPhD}, some also between planar electrodes with a
protrusion~\cite{Yi2002,TanjaProc}; here the protrusion is used as the inception point of the discharge.
Experiments of streamers between planar electrodes created with a laser were difficult to interpret
\cite{TanjaPhD}. The last two experiments were particularly designed to study streamers in homogeneous
background fields while the inception form curved electrodes is much easier. But even if one electrode is a
needle or a wire and the other one a plane, once the streamer tip approaches the planar electrode, the
background field is again well approximated by a homogeneous field. Finally, the electric field responsible
for sprite discharges, located between a charged thundercloud and the ionosphere, is rather homogeneous.

\subsection{Simulations and dependence on the initial ionization seed}

A streamer discharge in a homogeneous field is initiated by a localized ionization seed created, e.g., by a
cosmic particle shower or by an electric field inhomogeneity around a suspended particle. If the field is
above $\sim$30~kV/cm, a small seed first can undergo an avalanche phase where the ionization level increases
exponentially and then, once there are significant space charge effects, it reaches the streamer regime
\cite{Montijn2006/JPhD}. If the field is below threshold, a seed of finite size is required that after drift
separation of charges immediately enters the streamer phase.

\begin{figure}
\centering
\includegraphics[width=\figwidth]{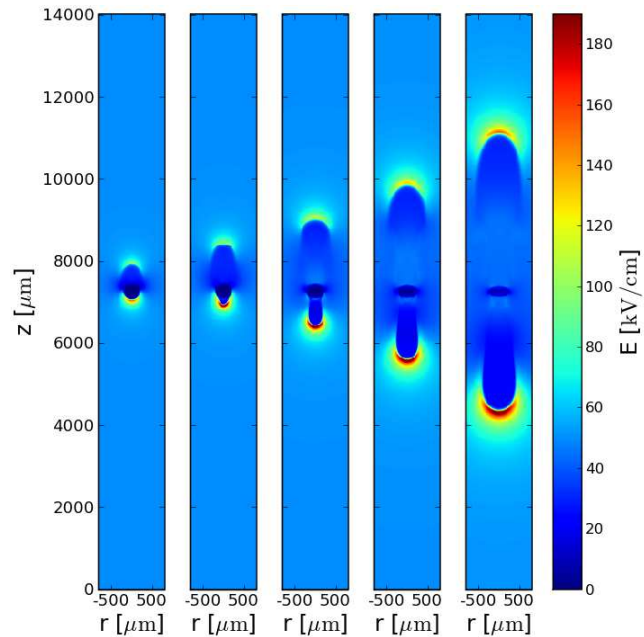}
\caption{Electric field of a double-headed streamer extending between two planar electrodes, plotted at
equal time steps $1.2 \units{ns}$.  The negative front is propagating upwards, the positive moves downwards.
The number of particles in the initial seed is $6\cdot10^{10}$.  Note that the lateral borders of the
figure do not correspond to the computational domain.
  }
  \label{double-head}
\end{figure}

\begin{figure}
a) \includegraphics[width=0.84\figwidth]{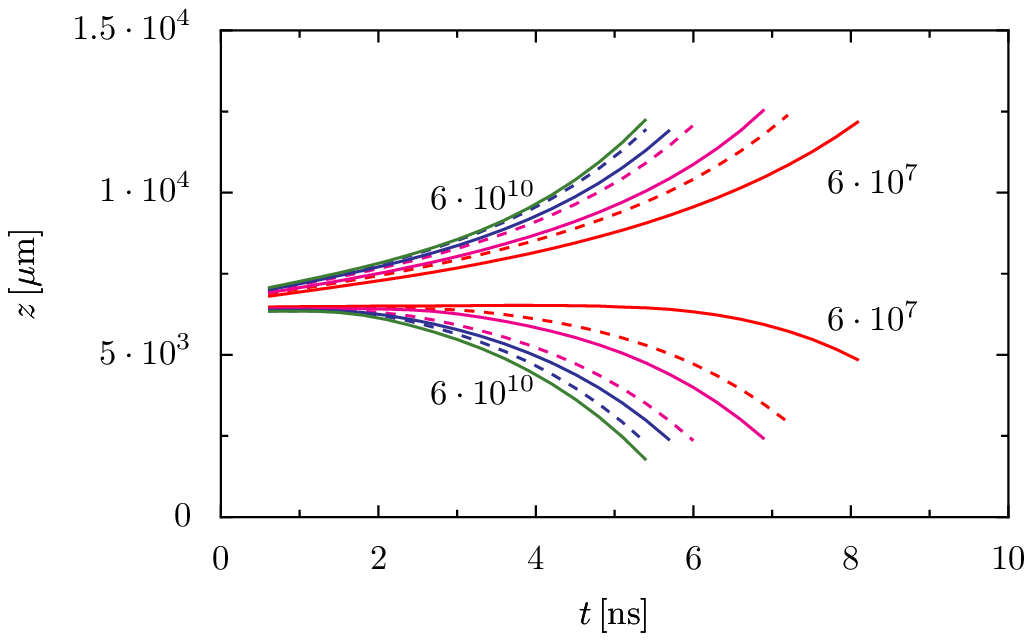} b)
\includegraphics[width=0.96\figwidth]{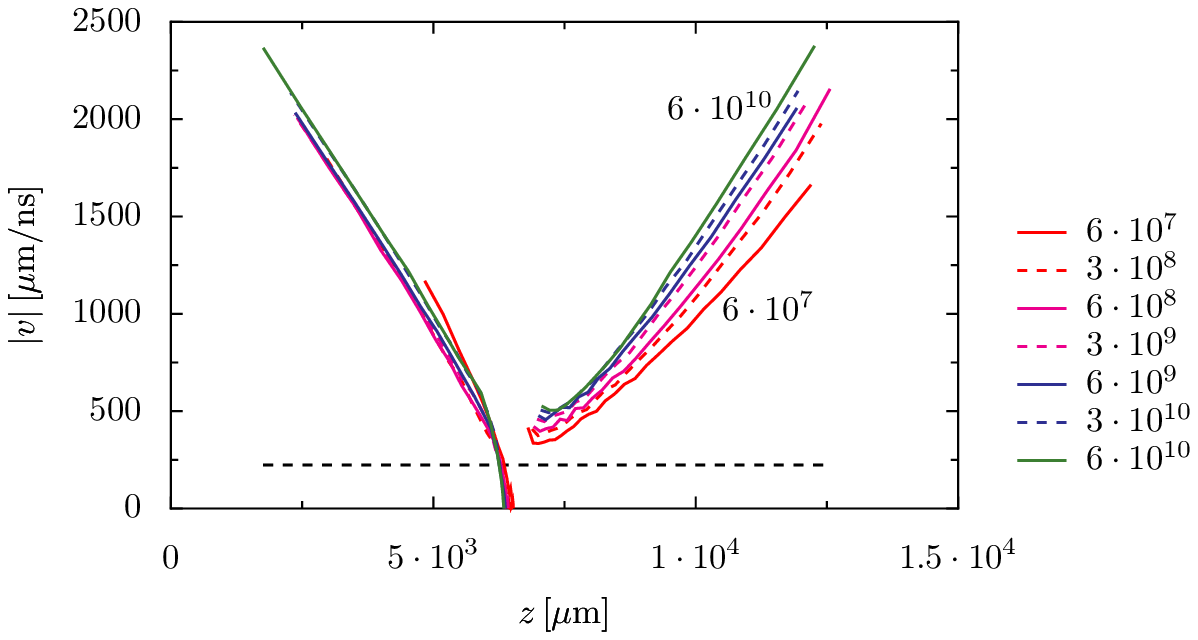}
\caption{a) Position $z(t)$ of the negative (upper) and
positive (lower) fronts of a double-headed streamer as a function of time $t$. The streamer evolves between
two planar electrodes with a background field $E_0 = 50 \units{kV/cm}$ as shown in Fig.~\ref{double-head}.
The distance between the electrodes is $14\units{mm}$.  The different colors correspond to different
numbers of initial electrons and $\mathrm{N}_2^+$ ions ranging
from $6\dexp{7}$ to $6\dexp{10}$ as indicated in the figure 
(the seed is initially neutral). b) The same data, now plotted as the absolute values of velocities $v(z)$ of the negative (right)
and positive (left) fronts as a function of the front location $z$. Note that each front propagates in a
different direction, but the absolute value of the velocity is plotted to help the comparison. The dotted
horizontal line indicates the velocity $v^*(E_0)$ of a planar negative ionization front in the background
field $E_0$. \label{posneg-hom-vels}}
\end{figure}

To investigate whether the initiation mechanism in a field above threshold has a lasting effect on the
propagation of a streamer, we run several simulations with different initial seeds.  The background field
was $E_0=50 \units{kV/cm}$ and the seeds had a spherical Gaussian profile with an $e$-folding radius of $74
\units{\mu m}$ located at the center of a $14 \units{mm}$-gap, creating a double-headed streamer. The
temporal evolution of the streamer shape is illustrated in Fig.~\ref{double-head} for a particular seed.
Panel a of Fig.~\ref{posneg-hom-vels} shows the position $z$ of the two ionization fronts as a function of
time $t$; it shows that both fronts accelerate in time, and that the fronts with the particle rich seeds are
ahead in evolution to those that start with a weaker seed; they propagate faster at any particular time both
on the positive and on the negative side.

The front velocity $v$ as a function of time $t$ is simply the derivative of the curves in panel a, and
$v(t)$ strongly depends on the size of the initial seed both for the positive and for the negative front. On
the other hand, panel b of Fig.~\ref{posneg-hom-vels} shows the absolute value of the velocity $v$ as a
function of position $z$ --- this is the observable typically measured in the experiments. Here the curves
$v(z)$ for the positive fronts essentially overlap for different seeds while those for the negative fronts
don't: the negative streamers with the largest seed have the largest velocities $v(z)$ as a function of head
position.

\subsection{Discussion of inception and propagation of positive and negative streamers}\label{diss3}

We now discuss the physical mechanisms causing the different dependencies of $v(z)$ on the initial seed. The
horizontal dotted line in panel b of Fig.~\ref{posneg-hom-vels} indicates the velocity $v^*(E_0)$ where
\begin{equation}\label{v*}
v^*(E)=|E|+2\sqrt{D\;|E|\;\alpha(|E|)}
\end{equation}
is the velocity of a planar negative ionization front in a field $E$ in the absence of photo-ionization in
dimensionless units~\cite{Ebert1997/PhRvE,Lagarkov1994}; it is given by the local electron drift velocity $|E|$ augmented
by the combined effect of electron diffusion $D$ and impact ion ionization $\alpha(|E|)$. The velocity
$v^*(E_0)$ evaluated in the background field $E_0$ is a lower bound for the velocity of a negative
ionization front with field enhancement and with photo-ionization as indeed can be seen in the figure.
Obviously, the negative front will always propagate at least with the electron drift velocity in the local
electric field. If the initial seed is stronger, field enhancement will build up faster while the front is
already in motion, and at each position $z$, the front is faster than for a weaker seed. The positive front,
on the other hand, has no lower bound for its velocity. The positive discharge side stays at rest until
photo-ionization has built up a sufficient electron concentration for the streamer to start. This means that
the inception time now strongly depends on the seed, but once the positive streamer propagates, it does it
with a similar velocity $v$ as function of the position, rather independently of the seed.

In view of the experimental results, a most interesting question is the comparison of the velocities of
positive and negative streamers. If we fix the time elapsed after the seed of the double-headed streamer is
created, the negative streamer is faster than the positive one.  If, on the other hand, one compares the
velocities at a fixed distance from the initial seed, the picture is different: for small distances, the
negative streamers are faster, but they are overcome by the positive ones at larger distances from the
initial seed. Note also that the differences between positive and negative streamers become the smaller the larger the initial number of particles.

The only other studies of double headed streamers with photo-ionization in a homogeneous electric field are
performed by Liu and Pasko~\cite{LiuP04,Pasko2007/PSST}; in particular, table 2
in~\cite{LiuP04} shows that the characteristics of streamer propagation, namely field enhancement in the
streamer head and field screening in the streamer interior are stronger on the positive streamer head, and
that the positive streamer is faster. This is found at air models applicable to a height of 0, 30 and 70 km
in the atmosphere.

The larger velocity of the positive streamers is surprising if one takes into account that for identical
field enhancement and identical electron distribution in the leading edge of the ionization front, the
negative front will always be faster~\cite{Ebert1997/PhRvE} as electron drift supports propagation of
negative fronts and acts against it for positive fronts. However, inspection of Fig.~\ref{double-head}
shows that the positive streamer is more focused and the field at its head is more enhanced. Ultimately,
electron drift leads to a ``dilution'' of head focusing and field enhancement in negative streamers and makes
them run slower at a given distance form the ionization seed.

\section{Streamers in needle-plane geometries}
\label{sect:inhomo}

The study of streamers disconnected from electrodes in homogeneous fields already gives a good qualitative
insight into many of their properties. However, laboratory experiments and engineering applications are
mostly done in inhomogeneous fields and streamers emerge from pointed electrodes. We therefore here study
positive and negative streamers emerging from a needle electrode and propagating towards a planar electrode. This
electrode configuration was implemented by means of the Charge Simulation Technique as described in
\ref{CST}.

\subsection{Weak dependence on the initial conditions}

\begin{figure}
\centering
\includegraphics[angle=0,width=\figwidth]{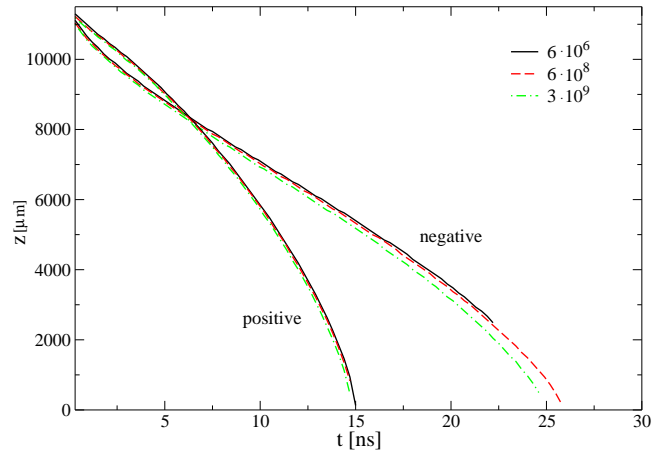}
\caption{Front propagation of negative and positive streamers for different seeds between needle-plane
electrodes. The gap is $L=11.5\,\rm{mm}$ long with an applied voltage of $23\units{kV}$; the needle parameters are $L_{needle}= 2.3\, \rm{mm}$, $R_{needle}=0.26\,\rm{mm}$. Different colours correspond to different number of particles in the initial seed.
 \label{posneg_zoft}}
\end{figure}

In order to study how the streamer behavior depends on parameters, we performed a number of different
simulations.  The first observation is that in contrast to the case of homogeneous fields, in inhomogeneous
fields the initial seed affects the propagation of the streamer only slightly, even when considering the
streamer velocity $v$ as a function of time $t$. This is illustrated in Fig.~\ref{posneg_zoft} where the
positions $z$ of streamers as a function of time $t$ are shown for several initial seeds --- this plot
corresponds to panel a in Fig.~\ref{posneg-hom-vels}.  The seeds are placed at the top of the gap and have spherical Gaussian profiles with radius $92\mu\rm m$ and between $6\dexp{6}$ and $3\dexp{9}$ particles.

 For positive
streamers in needle-plane geometries, this effect was found before by Pancheshnyi \emph{et al.}
\cite{Pancheshnyi2001/JPhD}. The reason is probably that in the high field region near the needle electrode,
the seed grows very rapidly compared to the slower evolution in the lower fields away from the needle.

\subsection{Simulations of positive streamers}

\begin{figure}
a) \includegraphics[width=0.9\figwidth]{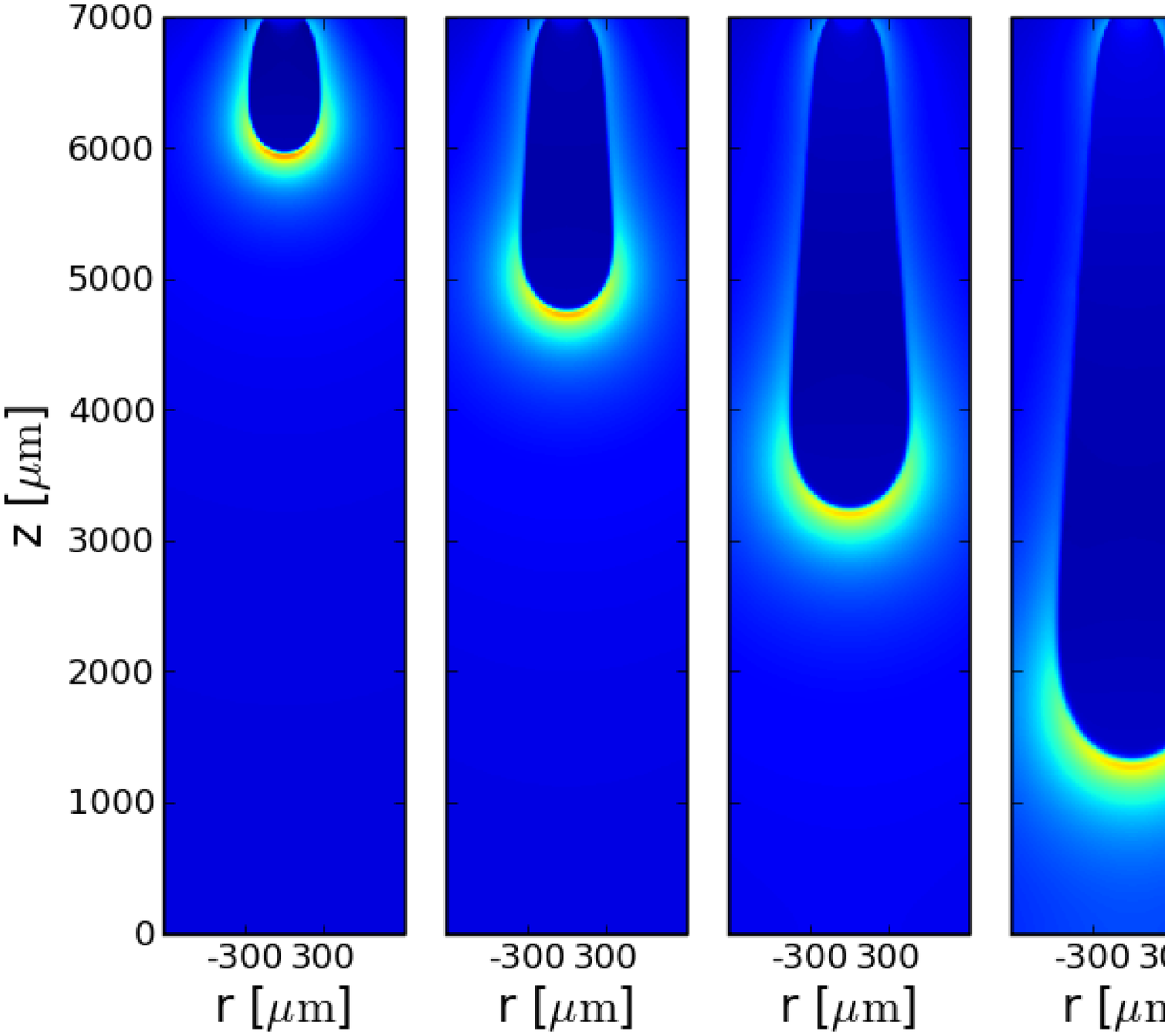} b)
\includegraphics[width=\figwidth]{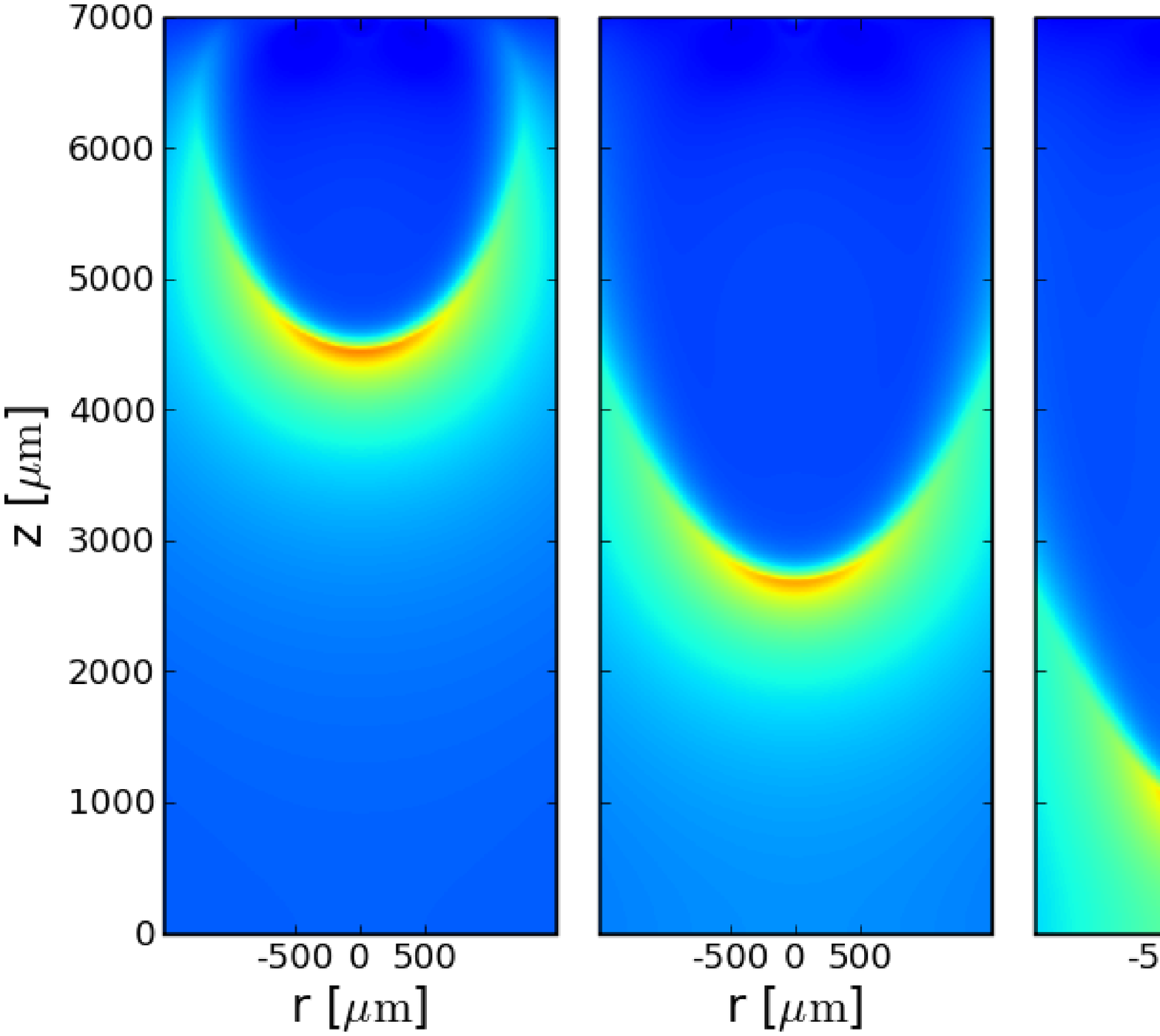}
\caption{Electric field created by a propagating streamer in a needle-plane geometry plotted at equal time
steps. Note that the computational domain extends beyond the lateral borders of the plot. a) A positive
streamer at time steps of $2.7\units{ns}$ in a voltage of $14\units{kV}$, b) a negative streamer at time steps of
$4.5\units{ns}$ also in a voltage of $14\units{kV}$. \label{inhom-eabss}}
\end{figure}

\begin{figure}
\includegraphics[width=2\figwidth]{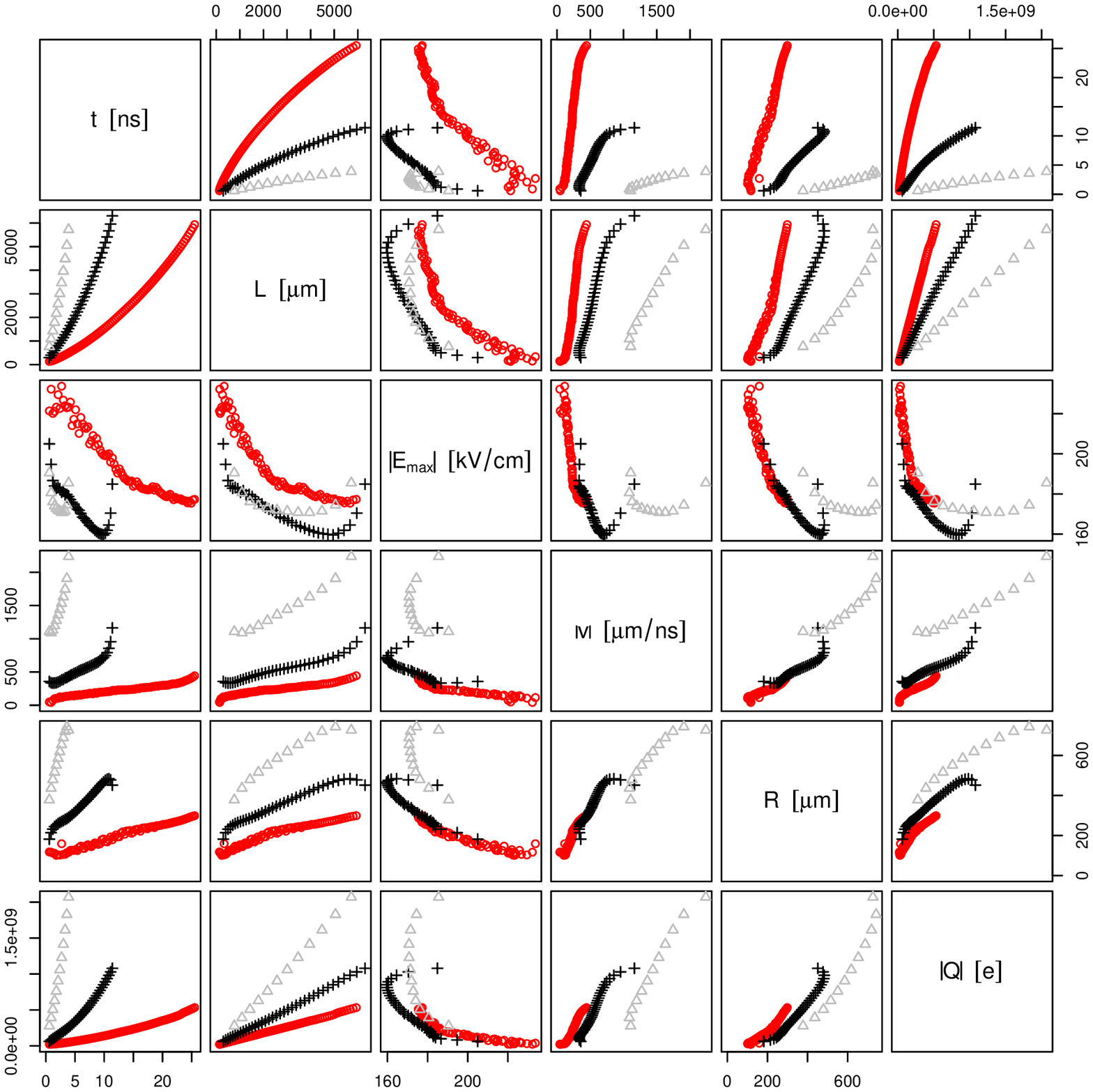}
\caption{\label{pairs-pos} Characterization of positive streamers by their main (electro-)dynamical
variables, namely time $t$, length $L$, maximal field $E_{max}$, velocity $v$, head radius $R$ and charge in
the streamer head $Q$; these quantities are defined in~\ref{defs}. The picture shows each variable
as a function of each other variable; the plot with variable $A$ in the $X$ axis and variable $B$ in the $Y$
axis is located in the column and row which have, respectively, $A$ and $B$ in their diagonal cells. Hence,
every figure appears twice with its axes swapped. Red circles correspond to a voltage of $10.5 \units{kV}$,
black crosses to $14\units{kV}$ and gray triangles to $21\units{kV}$. Note that all plots are qualitatively
similar for different applied voltages. Remarkably, the plots of streamer velocity $v$ as a function of
radius $R$ nearly fall on the same line for different voltages. 
}
\end{figure}

We now analyze positive streamers in the needle-plane geometry in more detail. We run simulations of
positive streamers in a short gap of $7\units{mm}$ gap at a voltage of $10.5 \units{kV}$, $14 \units{kV}$
and $21 \units{kV}$. The needle electrode had a radius $R_{needle} = 0.2 \units{mm}$ and a length
$L_{needle} = 2 \units{mm}$.  An example of the evolution of the streamers is shown in panel a of
Fig.~\ref{inhom-eabss}. An array plotting the relationship between each pair of the derived quantities
defined in \ref{defs} for each of our runs is provided in Fig.~\ref{pairs-pos}. The purpose of this array
representation is to find consistent relationships between the streamer characteristics. The overall
evolution shown in Fig.~\ref{pairs-pos} can be summarized as follows: as the streamer advances, it becomes
thicker and faster and the total charge in the streamer head increases while the enhanced electric field
decreases.

\subsection{The velocity of positive streamers}\label{vplus}

\begin{figure}
\centering
\includegraphics[width=\figwidth]{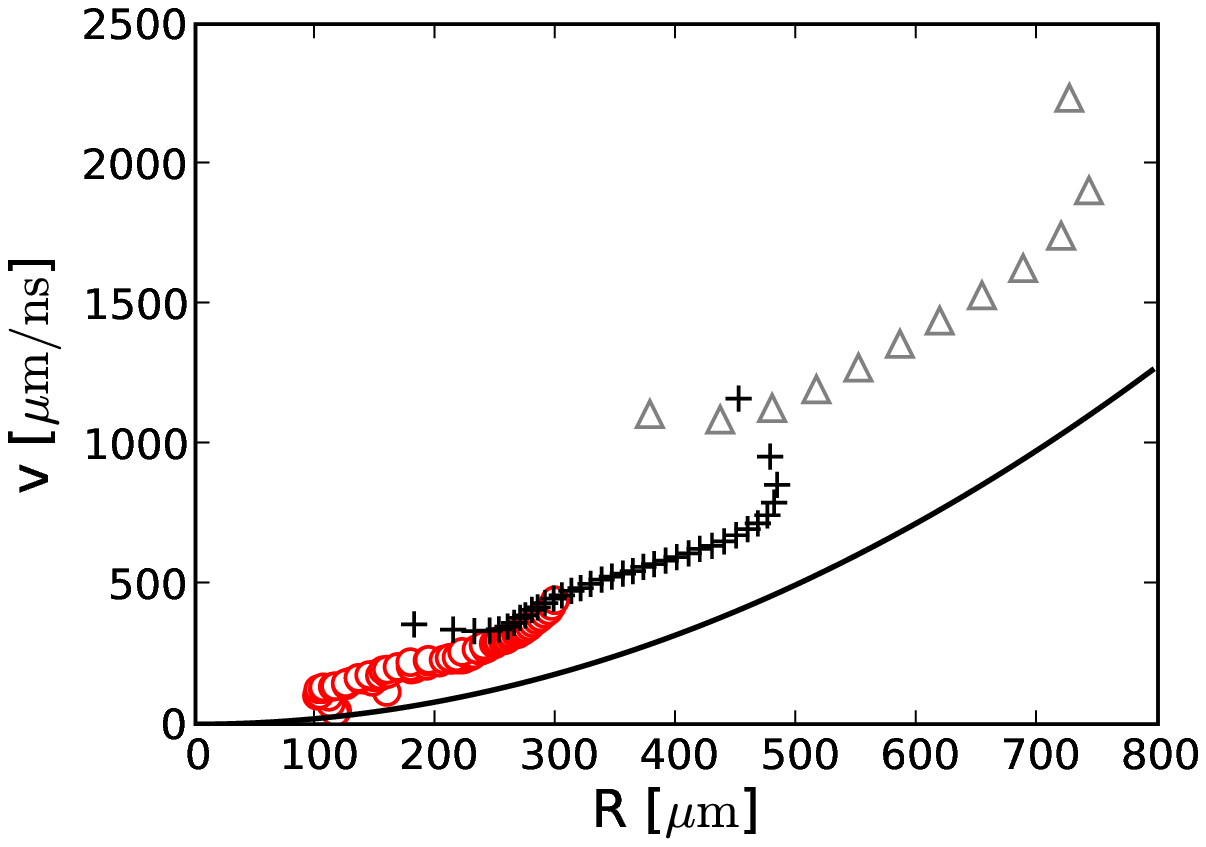}
\caption{Relationship between velocity $v$ and radius $R$ of positive streamers.  The symbols are simulation
results as presented in Fig.~\ref{pairs-pos}, the continuous line represents the empirical fit
\parref{fittanja}.
  \label{theocomp-rv}}
\end{figure}

While for a negative streamer, the velocity $v$ cannot become smaller than the electron drift velocity
$|E_{max}|$ in the locally enhanced field, for the positive streamer the velocity $v$ increases while the
field enhancement $|E_{max}|$ decreases. In the experimental investigation~\cite{BrielsPM}, the completely
empirical relation
\begin{equation}
  v \approx \frac{0.5~d^2}{\units{mm~ns}}
  \label{fittanja}
\end{equation}
between velocity $v$ and diameter $d$ of positive streamers in air at standard temperature and pressure was
fitted to the experimental data (in Fig.~6b of~\cite{BrielsPM}). In Fig.~\ref{pairs-pos}, the plots of $v$
as a function of radius $R$ also lie more or less on one line. In Fig.~\ref{theocomp-rv} we therefore
compare the $v(R)$ plot of our simulation results with the empirical fit (\ref{fittanja}) to the
experiments. The plot does not contain any fitting parameters and shows a close resemblance between
experiment and simulation. When interpreting the figure, it should be noted that the simulations measure the
\emph{geometrical} radius of the space-charge layer, also called ``electro-dynamic radius'', while an
experiment measures the visible, or radiative, radius. There can be a significant difference between both
measures;  in \cite{Pancheshnyi2005/PhRvE} it is estimated that the electro-dynamic radius is about twice the
radiative radius.

A positive streamer propagates due to the photo-ionization in front of its head. Comparing the head radius
to the photo-ionization lengths~\cite{Luque2007/arXiv}, the hypothesis of Kulikovskii~\cite{KuliHead} that
the streamer radius would be determined by the photo-ionization length can be clearly discarded. The
photo-ionization absorption is fitted by two lengths scales, one of them is essentially negligibly small,
the other one much larger than the head radius~\cite{Luque2007/ApPhL,Luque2007/arXiv}. Free electrons
created by photo-ionization are therefore available throughout the region where the electric field is above
the ionization threshold. Therefore the streamer velocity will be mainly determined by the size of the
region where further electron multiplication by impact is efficient. This region, in turn, is roughly
determined by two factors: the enhanced electric field and the electro-dynamic radius of the streamer. In
Fig.~\ref{pairs-pos} we see that the radius varies much more than the enhanced field, which explains the
qualitative relation between velocity and radius.

\subsection{Simulations of negative streamers}

We investigated negative streamers also in a $7 \units{mm}$-gap with applied voltages of $10.5\units{kV}$
and $14\units{kV}$. The electrode geometry was the same as for positive streamers. The evolution of the
spatial structure is illustrated in panel b of Fig.~\ref{inhom-eabss}. As in the case of the homogeneous
field in Fig.~\ref{double-head}, the negative streamer is broader and the field is less enhanced.

\begin{figure}
\centering
\includegraphics[width=2.0\figwidth]{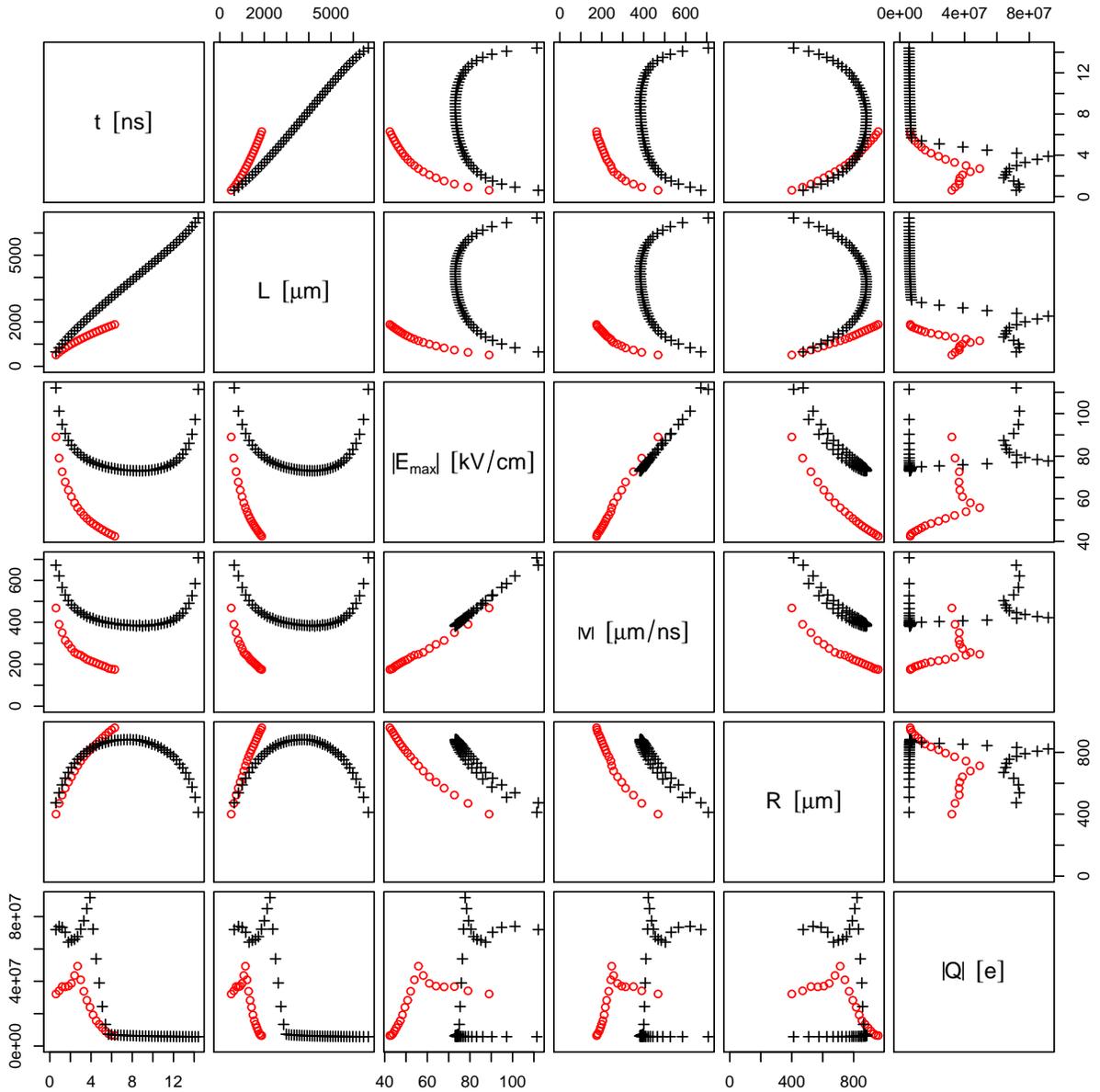}
\caption{\label{pairs-neg} Characterization of negative streamers by their main (electro-)dynamical
variables, the presentation is the same as in Fig.~\ref{pairs-pos} for positive streamers. All simulations
were performed in the same geometry as for the positive streamers, with applied voltages $10.5\units{kV}$
(red circles) and $14 \units{kV}$ (black crosses). Note that in the low voltage case, the enhanced field
ceases to be strong enough to sustain the streamer propagation.}
\end{figure}

Figure~\ref{pairs-neg} shows the relationship between relevant (electro-)dynamic quantities during the
evolution of these two streamer simulations. The streamer in the lower voltage (red symbols) becomes slower,
the enhanced field eventually drops below the threshold for impact ionization, the head charge disappears
and the radius diverges. At that instant of time, the remaining electrons from the streamer head continue to
drift towards the planar electrode, but the impact ionization ceases to be efficient and the streamer mode
of propagation stops. This is probably the generic way how a negative streamer extinguishes, quite
different form the one reported for positive streamers in~\cite{PanchCease}. The streamer in the higher
voltage undergoes a similar intermediate evolution. However, eventually the proximity of the planar
electrode again enhances the field and the streamer reaches the electrode.

\subsection{Velocity of negative streamers}

As already discussed above and in section~\ref{diss3}, negative streamers propagate not only due to
photo-ionization but also due to electron drift. For the velocity, this implies a stronger dependence on the
enhanced field, that thus overcomes the dependence on the radius. In fact, comparison of
Figs.~\ref{pairs-pos} and \ref{pairs-neg} shows that the velocity of positive streamers increases with
radius while the field enhancement decreases; the velocity of negative streamers, on the other hand, increases with field
enhancement while the radius decreases. Indeed, Fig.~\ref{pairs-neg} shows a very clear correlation between
velocity $v$ and field enhancement $E_{max}$ for negative streamers that we now analyze further.

\begin{figure}
\centering
\includegraphics[width=\figwidth]{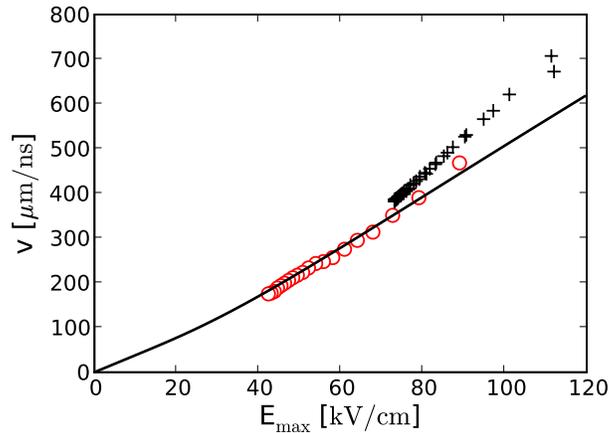}
\caption{Relationship between velocity $v$ and enhanced field $E_{max}$ for negative streamers. The symbols
are simulation results as presented in Fig.~\ref{pairs-neg}, the continuous line represents the velocity of
a planar front without photo-ionization $v^*(E_{max})$ (\ref{v*}).
  \label{theocomp-ev}}
\end{figure}

Actually, one can compare the actual velocities with the velocity $v^*(E_{max})$ where $v^*(E)$ is given in
Eq.~(\ref{v*}). The velocity $v^*(E_{max})$ is the
velocity of a planar fully relaxed negative streamer front in a field $E_{max}$ when the effect of
photo-ionization is neglected. Fig.~\ref{theocomp-ev} shows the simulation data for $v(E_{max})$ from
Fig.~\ref{pairs-neg} and the function $v^*(E_{max})$ for comparison. The coincidence is strong. Deviations
mainly come from the fact that the analytical equation is a lower bound to the actual velocity as
photo-ionization is neglected (for a comparison of simulation data without photo-ionization with the
analytical formula we refer to~\cite{Brau2008/PhRvE,Brau08}).

One should remark here that the background electric field under which the streamer propagates most of the
time is quite low and hence one does not observe a buildup of the ionization level in front of the streamer
as in the high-field case of \cite{Luque2007/ApPhL}.  Therefore, we do not observe a transition to a regime
dominated by photo-ionization.

\subsection{Comparison of positive and negative streamer simulations}

\begin{figure}
\centering
\includegraphics[angle=0,width=\figwidth]{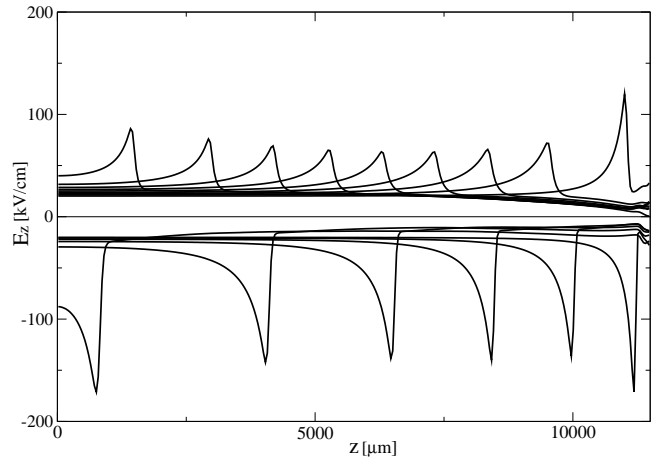}
\caption{Electric field on the streamer axis at equal time steps of $3\units{ns}$ for negative (upper) and positive
(lower) streamers propagating to the left. 
The number of particles in the initial seed is $6\dexp{8}$.
} \label{posneg_eofz}
\end{figure}

We here directly compare the propagation of a positive and a negative streamer in a longer gap of 11.5~mm
than considered above in needle-plane geometry.  For both positive and negative streamers the applied voltage is $V=23\,\rm kV$. The radius of the needle is $R_{needle}=0.26\, \rm mm$ and its length is $L_{needle}=2.3\,\rm mm.$ The position as a function of time was already presented in~Fig.~\ref{posneg_zoft} where it was shown that the
evolution depends only very weakly on the initial ionization seed. The figure shows that the negative
streamer initially is faster, but it is soon overtaken by the positive one.

In Fig.~\ref{posneg_eofz} we show the spatial profiles of the electric field on the streamer axis for a
number of time steps; the streamers are propagating to the left. The electric field at the positive streamer
heads is much more enhanced than on the negative ones. This larger field enhancement is due to the smaller
radii of the positive streamers that consecutively propagate much faster. Also the field inside the streamer
channel is screened less for negative streamers. We note that in the only other comparable simulation of
positive and negative streamers by Babaeva and Naidis~\cite{Babaeva1997/ITPS}, the field inside the negative
streamer channel is higher as well, but Fig.~6 of that paper shows that their negative streamers are faster
than their positive ones, though there is also a consistency problem between their Figs.~6 and~7, and
positive and negative streamer velocities are not compared in the text.

\section{Comparison with experiments}\label{exp}

Experiments~\cite{Winands2008/JPhD,BrielsPM} show that positive and negative streamers in ambient air driven
by voltages above 60~kV behave qualitatively similar. On the other hand, there are major differences below
40~kV~\cite{BrielsPM}. Positive streamers form at lower applied voltages, they are faster, longer and
thinner. Our simulations at voltages between 10.5 and 21~kV in shorter gaps reproduce all these features.

\subsection{Inception}

While the full inception process in interaction with the electrode needle surface is not part of the present
simulations, we observe that positive streamer inception is not very sensitive to the initial ionization
seed while the negative streamer formation depends on it, at least in homogeneous fields.

\subsection{Velocity}

Close to the needle electrode, the electrons of the negative ionization seed drift outwards in the local
field and are faster. However, just the lack of outward drift motion in the positive seed leads eventually
to a larger field enhancement and ultimately to a faster propagation of the positive streamer at the same
distance from the electrode.

Experimental measurements of the velocity of positive streamers are very well fitted by the empirical
equation~(\ref{fittanja}) that relates their velocity to their radius. Figure~\ref{theocomp-rv} shows that
this equation also fits our simulation results quite well without any fit parameter.

\subsection{Diameter}

The negative streamers are thicker and less focused, both in simulations and in experiment. The minimal
diameter of positive streamers in our simulations is about 0.2~mm, identical to the minimal diameter
reported in experiments~\cite{Briels2006/JPhD,BrielsScaling,BrielsPM}. It should be noted, however, that the
definition of the radius might differ between experiments and simulations as discussed in
section~\ref{vplus}.

\subsection{Length and extinction}

In a potential of 10.5~kV, the negative streamers extinguish after less than 2 mm, while the positive ones
reach the planar electrode at 7 mm distance. In fact, in the experimental paper~\cite{BrielsPM}, discharges
of 2 mm length are not called streamers, and the extinction of these very short negative discharges is in
agreement with experiment. At 14~kV, our simulated negative streamers do reach the planar electrode, but
they are helped by a strong initial ionization seed and a short gap. Simulations in longer gaps are in
progress.

\section{Conclusion and outlook} \label{sect:summary}

We have studied the propagation of double-headed streamers in a homogeneous field and the inception and
propagation of positive and negative streamers emerging from needle electrodes. We have shown that for
spatially concentrated ionization seeds containing 
from about $10^7$ to about $10^{10}$ 
electron-ion pairs, the streamer velocity at a
given streamer length depends only weakly on the seed, except for the case of negative streamers in
homogeneous fields. We have found qualitative and quantitative agreement with experiment as summarized in
section~\ref{exp}.

We have shown that the relations between velocity $v$, radius $R$, field enhancement $E_{max}$ and head
charge $Q$ that characterize the propagation of a streamer, differ qualitatively between streamers of
different polarity. The velocity of a positive streamer in air is mainly determined by its radius, in
accordance with the empirical fit formula~(\ref{fittanja}), while that of a negative one is dominated by the
enhanced electric field and well approximated by $v^*(E_{max})$ where $v^*(E)$ is the velocity of a planar
negative ionization front in the absence of photo-ionization. These statements on the simulations hold in
the voltage range of 10 to 20~kV in gaps of 7 or 11.5~mm length. Longer gaps and higher potentials will be
investigated in the future.

The characterization of the streamer head by velocity, radius, field and charge is a first step towards an
electro-dynamical characterization of streamer head and channel. Such models were already sketched
in~\cite{DyaKach1,DyaKach2,RaizerSim,Baz}, however, they need to take care of the polarity dependence of
streamers, and they have to be made consistent with charge conservation~\cite{Ebert2006/PSST}.

The fact that in experiments~\cite{BrielsPM} positive streamers moved faster than negative ones, initially
was quite puzzling from a theoretical point of view. As electron drift acts against positive and in favor of
negative streamers, a streamer with identically formed space charge layer in its head will always move
faster, if it is negative. However, the simulations show that the electron drift does not necessarily help
the negative streamer to propagate; rather the outwards drift motion that is essentially linear in the
field, leads to a growth of the head radius and a subsequent ``dilution'' of field enhancement. The growth of
the positive streamer depends more non-linearly on the local field through impact ionization; therefore it
stays thinner, the field is enhanced more and subsequently it propagates faster. In experiments, this
asymmetry is found to decrease with increasing voltage; whether simulations show the same, will also have to
be investigated in the future.

Streamer physics is an exciting and widely open field, and it is amazing to note how little their polarity
dependence has been characterized up to now, both experimentally and theoretically. Of course, the study of
single streamers is only one problem in a range of phenomena, other questions concern streamer
branching~\cite{Montijn2006/PRE,Nijdam2008/APL}, interactions~\cite{Luque2007/arXiv,Brau08}, particle
aspects~\cite{Li07,Li08}, the full inception process near an electrode or the electro-dynamical
characterization of multi-streamer processes.


\appendix

\section{Simulation of a needle electrode}
\label{CST} 
\begin{figure}
\centering
\includegraphics[width=\figwidth]{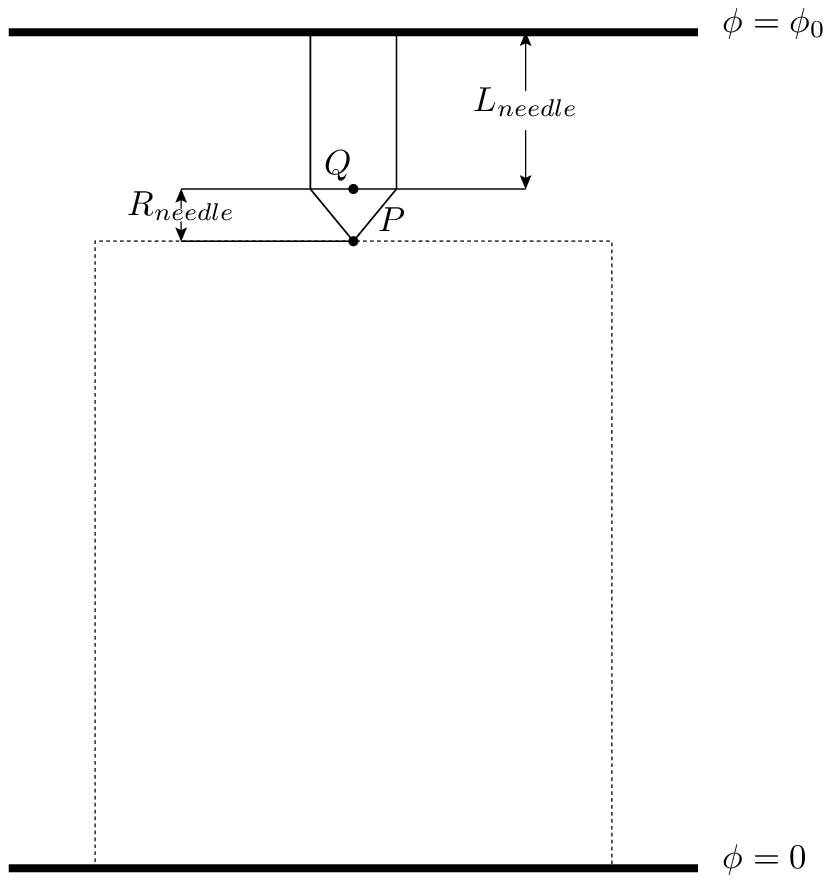}
\caption{Implementation of a needle electrode using the Charge Simulation
Technique (CST) with a single charge.  The particles are restricted to the
rectangular domain below the needle, while the electrostatic field is solved
in a larger domain limited by the two planar electrodes.
  \label{schema}}
\end{figure}

We simulated needle electrodes by a simplified version of the Charge Simulation Technique (CST)
described in \cite{Singer1976/TPAS}. In general, the presence of an electrode imposes a fixed electrostatic
potential along its surface.  However, one can retain the main properties of a needle electrode by fixing
the potential only at its tip (point $P$ in Fig.~\ref{schema}).  This is achieved by introducing a simulated
point charge $Q$ at a certain location inside the electrode.  At each time step of the simulation, the value
of $Q$ is calculated to keep $\phi(P)$ (the electrostatic potential at $P$) fixed to $\phi(P)=V_0$. This
schematic approach approximates the effect of a needle with a radius equal to the distance between $P$ and
$Q$ ($R_{needle}$ in Fig.~\ref{schema}) and a length equal to the distance between the upper planar
electrode and $Q$ ($L_{needle}$ in Fig.~\ref{schema}).

In our simulations, we restrict the particles to the cylindrical volume below the needle tip and apply a
homogeneous Neumann boundary condition at its top and bottom sides. Although this creates an artificial
boundary inside the physical domain, note that our streamers will touch this plane only around the needle
tip.  Hence it can be used as a rough approximation for an electrode with a free in- or outflow of
electrons.

\section{Electro-dynamic characterization of the streamer}
\label{defs}

From one numerical simulation, one gets spatial profiles of the densities of species and electric fields.
However, to help the physical understanding of the streamer process it is often useful to use more
macroscopic quantities that roughly characterize the propagation of a streamer.  Although the meaning of
quantities such as radius or length is intuitively obvious, there is a certain unavoidable arbitrariness in
the way one defines them from the more microscopic data.  Therefore we provide here a precise description of
the way we define each of these quantities.

First of all we define the streamer tip as the point on the propagation axis where the absolute value of
charge density is maximal.  The space-charge layer of the streamer is then defined as the volume around the
streamer tip where the absolute value of the charge density is larger than half of its maximum value. The
streamer head charge $Q$ is defined as the net charge content of the space-charge layer.

The streamer length $L$ is the separation between the needle electrode and the streamer tip.  The streamer
velocity is simply $v = dL/dt$. The enhanced field $E_{max}$ is defined as the maximum of the electric field
in absolute value.

The definition of a streamer radius is somewhat more involved.  We are mainly interested in characterizing
the shape of the space-charge layer.  Hence we followed this procedure: for each $z$ we took the radius $r$
with the highest charge density.  This gives us a $r(z)$ curve that we restrict to the points inside the
space-charge layer. This curve is fitted to a circle and we take the resulting radius as the radius of the
streamer, $R$.

\ack A.L. is financially supported by STW under contract number 06501, and V.R. by IOP-EMVT under contract
number 062126B.


\section*{References}

\begin{thebibliography}{18}

\bibitem{Raizer91} Raizer Yu P 1991, {\em Gas Discharge Physics}, Berlin: Springer

\bibitem{BrielsPM} Briels T M P, Kos J, Winands G J J, van Veldhuizen E M and Ebert U 2008,
    {\em Positive and negative streamers in ambient air: measuring diameter, velocity and dissipated energy},
    submitted to the special issue on "Streamers, Sprites and Lightning" in \JPD

\bibitem{Baz} Bazelyan E M, Raizer Yu P 1998, {\em Spark Discharge} (CRC Press, Boca Raton, Florida)

\bibitem{Pasko2002/GeoRL} Pasko V P, Stenbaek-Nielsen H-C 2002, {\em Geophys. Res. Lett.} {\bf 29} 82

\bibitem{Williams06/PSST} Williams E R 2006, \PSST {\bf 15} S91

\bibitem{Winands2006/JPhD} Winands G J J, Liu Z, Pemen A J M, van Heesch E J M, Yan K, van Veldhuizen E M
2006, \JPD {\bf 39} 3010

\bibitem{Winands2008/JPhD} Winands G J J, Liu Z, Pemen A J M, van Heesch E J M, Yan K 2008,
    accepted for the special issue on "Streamers, Sprites and Lightning" in \JPD

\bibitem{win07} Winands G J J 2007, {\em Efficient streamer plasma generation}, PhD thesis, Eindhoven
    University of Technology, The Netherlands. Available on www.tue.nl/bib.

\bibitem{Kunhardt} Wang M C, Kunhardt E E 1990, {\em Phys. Rev. A}
{\bf 42}, 2366

\bibitem{Dhali1987/JAP} Dhali S K, Williams P F 1987, {\em J. Appl. Phys.} {\bf 62} 4696

\bibitem{Kossyi1992/PSST} Kossyi I A, Kostinsky A Y, Matveyev A A, Silakov V P 1992, \PSST {\bf 1} 207

\bibitem{Babaeva1996/JPhD} Babaeva N Y, Naidis G V 1996, \JPD {\bf 29}, 2423

\bibitem{Abdel-Salam2007/JPhD} Abdel-Salam M, Nakano M, Mizuno A 2007, \JPD {\bf 40} 3363

\bibitem{Bourdon2007/PSST} Bourdon A, Pasko V P, Liu N Y, C{\'e}lestin S, S{\'e}gur P, Marode E 2007,
\PSST {\bf 16}, 656

\bibitem{Luque2007/ApPhL} Luque A, Ebert U, Montijn C, Hundsdorfer W 2007, {\em Appl. Phys. Lett.} {\bf 90}, 1501

\bibitem{Pancheshnyi2005/PhRvE} Pancheshnyi S, Nudnova M, Starikovskii A 2005, {\em Phys. Rev E} {\bf 71}, 016407

\bibitem{Pancheshnyi2005/PSST} Pancheshnyi S 2005, \PSST {\bf 14}, 645

\bibitem{Yi2002} Yi W J, Williams P F 2002, \JPD {\bf 35} 205

\bibitem{Pancheshnyi2003/JPhD} Pancheshnyi S V, Starikovskii A Y,
  \JPD {\bf 36} 2692

\bibitem{Kuli1994} Kulikovsky A A 1994, \JPD {\bf 27} 2564

\bibitem{Babaeva1997/ITPS} Babaeva N Y, Naidis G V 1997, {\em IEEE Trans. Plasma Sci.} {\bf 25} 375

\bibitem{LiuP04} Liu N, Pasko V P 2004, {\em J. Geophys. Res.} {\bf 109} A04301

\bibitem{Pasko2007/PSST} Pasko VP 2007, \PSST {\bf 16}, 13

\bibitem{Luque2007/arXiv} Luque A, Ebert U, Hundsdorfer W 2007, {\em ArXiv.org e-prints:} 0712.2774.

\bibitem{Montijn2006/JCoPh} Montijn C, Hundsdorfer W, Ebert U 2006, {\em J. Comput. Phys.} {\bf 219}, 801

\bibitem{DyaKach1} D'yakonov  M  I and Kachorovskii V Yu 1988 {\em Sov. Phys. JETP} {\bf 67} 1049

\bibitem{DyaKach2} D'yakonov  M  I and Kachorovskii V Yu 1989 {\em Sov. Phys. JETP} {\bf 68} 1070

\bibitem{RaizerSim} Raizer Yu P and Simakov A N 1998 {\em Plasma Phys. Rep.} {\bf 24} 700

\bibitem{Ebert2006/PSST} Ebert U, Montijn C, Briels T M P, Hundsdorfer W, Meulenbroek B, Rocco A and
    van Veldhuizen E M 2006, \PSST {\bf 15} S118

\bibitem{Briels2006/JPhD} Briels T M P, Kos J, van Veldhuizen E M and Ebert U 2006, \JPD {\bf 39} 5201

\bibitem{BrielsScaling} Briels T M P, van Veldhuizen E M and Ebert U 2008, {\em Positive streamers in air
    and nitrogen of varying density: experiments on similarity laws},
    submitted to the special issue on "Streamers, Sprites and Lightning" in \JPD

\bibitem{TanjaProc} van Veldhuizen E M, Rutgers W R, Ebert U 2002, {\em Branching of streamer type
    corona discharge}, {\em proceedings of the XIV Int. Conf. Gas Discharges and Appl.} (Liverpool, UK)

\bibitem{TanjaPhD} Briels T M P 2007, {\em Exploring streamer variability in experiments} (Ph.D. thesis,
    Eindhoven Univ. Techn., The Neterlands)

\bibitem{Montijn2006/JPhD} Montijn C, Ebert U 2006, \JPD {\bf 39} 2979

\bibitem{Lagarkov1994} Lagarkov A N, Rutkevich I M, {\em Ionization Waves in Electrical Breakdown in Gases}, (New York: Springer-Verlag), 1994 

\bibitem{Ebert1997/PhRvE} Ebert U, van Saarloos W, Caroli C 1997, {\em Phys. Rev. E} {\bf 55} 1530

\bibitem{Pancheshnyi2001/JPhD} Pancheshnyi S V, Starikovskii A Y 2001, \JPD {\bf 34} 248

\bibitem{KuliHead} Kulikovsky A A, 2000, \JPD {\bf 33} 1514

\bibitem{PanchCease} Pancheshnyi S, Starikovskii A Y 2004, \PSST {\bf 13} B1

\bibitem{Brau2008/PhRvE} Brau F, Luque A, Meulenbroek B, Ebert U, Sch{\"a}fer  L 2008,
  {\em Phys. Rev. E} {\bf 77}, 026219

\bibitem{Brau08} Luque A, Brau F, Ebert U 2007, {\em ArXiv.org e-prints:} 0708.1722

\bibitem{Montijn2006/PRE} Montijn C, Ebert U, Hundsdorfer W 2006, {\em Phys. Rev. E} {\bf 73} 065401

\bibitem{Nijdam2008/APL} Nijdam S, Moerman J S, Briels T M P, van Veldhuizen E M, Ebert U 2008,
    {\em Appl. Phys. Lett.} {\bf 92} 101502

\bibitem{Li07} Li C, Brok W J M, Ebert U, van der Mullen J J A M 2007, {\em J. Appl. Phys.} {\bf 101} 123305

\bibitem{Li08} Li C, Ebert U, Brok W J M, Hundsdorfer W 2008, \JPD {\bf 41} 032005

\bibitem{Singer1976/TPAS} Singer H, Steinbigler H, Weiss P 1976, {\em IEEE Trans. Power App. and Sys.}
{\bf 93} 1660

\end{thebibliography}
\end{document}